\documentclass{article}
\usepackage{geometry}
\usepackage{graphicx}
\usepackage{cite}
\usepackage{amsmath}
\usepackage[breaklinks=true,colorlinks=true,ps2pdf,urlcolor=blue]{hyperref}

\usepackage{floatrow}
\newfloatcommand{capbtabbox}{table}[][\FBwidth]

\usepackage{subfigure}

\usepackage{tkz-linknodes}

\begin{document} 
\title{Reflective Arrayed Waveguide Grating with Sagnac Loop Reflectors in Silicon-on-Insulator with Gaussian Pass-band}
\author{B.~Gargallo$^1$, P.~Mu\~noz$^{1,2}$, R. Ba\~nos$^1$, A.L. Giesecke$^3$,\\ J. Bolten$^3$, T. Wahlbrink$^3$ and H.~Kleinjans$^3$\\
$^1$Universitat Polit\`ecnica de Val\`encia \& $^2$VLC Photonics S.L.\\ c/ Camino de Vera s/n - 46022 Spain \\ \href{mailto:pmunoz@iteam.upv.es}{pmunoz@iteam.upv.es} \\
$^3$AMO GmbH \\ Otto-Blumenthal-Stra\ss e 25, 52074 Aachen, Germany \\ \href{mailto:kleinjans@amo.de}{kleinjans@amo.de}}

\maketitle

\begin{abstract}
In this paper the experimental demonstration of a Silicon-on-Insulator Reflective Arrayed Waveguide Grating (R-AWG) is reported. The device employs one Sagnac loop mirror per arm in the array, built with a 1x2 input/outputs, 50:50 splitting ratio, Multimode Interference coupler, for total reflection. The spectral responses obtained are compared to those of regular AWGs fabricated in the same die.
\end{abstract}

\section{Introduction}
Wavelength multi/demultiplexers are central components in optical telecommunication networks, subject to demanding requirements, both in terms of performance and manufacturing. Photonic integration is usually the basis for large count WDM multiplexers. The cost of an integrated circuit is fundamentally related to its footprint \cite{kirchain_nphot}. In terms of footprint, reflective multiplexers as the Echelle Diffraction Grating (EDG) achieve considerable size reduction \cite{lycett}. One issue with EDGs is to maximize the reflection on the grating, in order to minimize the overall insertion losses, with the deposition of metal layers at the edge of the grating \cite{feng}, or the addition to the grating of other structures such as Bragg reflectors \cite{ryckeboer}. While the deposition of metals supplies broadband reflectors, it requires resorting to additional fabrication steps. Conversely, Bragg reflectors can be manufactured in the same steps than the EDG, but it is well known the reflection bandwidth is inversely proportional to their strength \cite{pruessner}. Similarly, AWG layouts with reflective structures midway in the array, i.e. reflective AWGs (R-AWG) are possible as well. The reflectors can be implemented in analog ways to the ones for the EDGs, as reflective coatings on a facet of the chip \cite{inoue, soole}, photonic crystals \cite{dai}, external reflectors \cite{peralta2} and Bragg reflectors \cite{okamoto_dbr}. A common issue of all the described approaches for the reflector is that broadband full reflectivity requires additional fabrication steps, and therefore increases the final cost of the multiplexer. In \cite{ikuma_ecoc2007} a configuration for a R-AWG, where the well known Sagnac Loop Reflectors (SLR) are used as reflective elements at the end of the arrayed waveguides, was demonstrated in Silica technology. A SLR is composed of an optical coupler with two output waveguides, that are connected to each other forming a loop. These reflectors are broadband, can supply total reflection, and can be fabricated in the same lithographic process than the rest of the AWG. Moreover, the reflection of a SLR depends on the coupling constant of the coupler. Hence, it can be different for each of the waveguides in the array. The modification of the field pattern in the arrayed waveguides of an AWG allows for spectral response shaping, as for example box like transfer function \cite{okamoto} and multi-channel coherent operations \cite{doerr} amongst other. In this paper we report on the experimental demonstration of a Silicon-on-Insulator R-AWG based on SLRs with Gaussian pass-band response. This opens the door to further research on non Gaussian response R-AWGs, using MMIs with coupling ratios other than 50:50 as we propose in \cite{gargallo_arxiv_rawg_model}.

\section{Design, fabrication and characterization}
\begin{figure}
  {\par\centering
  \subfigure[Optical microscope image of the fabricated devices]{\resizebox*{0.43\textwidth}{!}{\includegraphics*{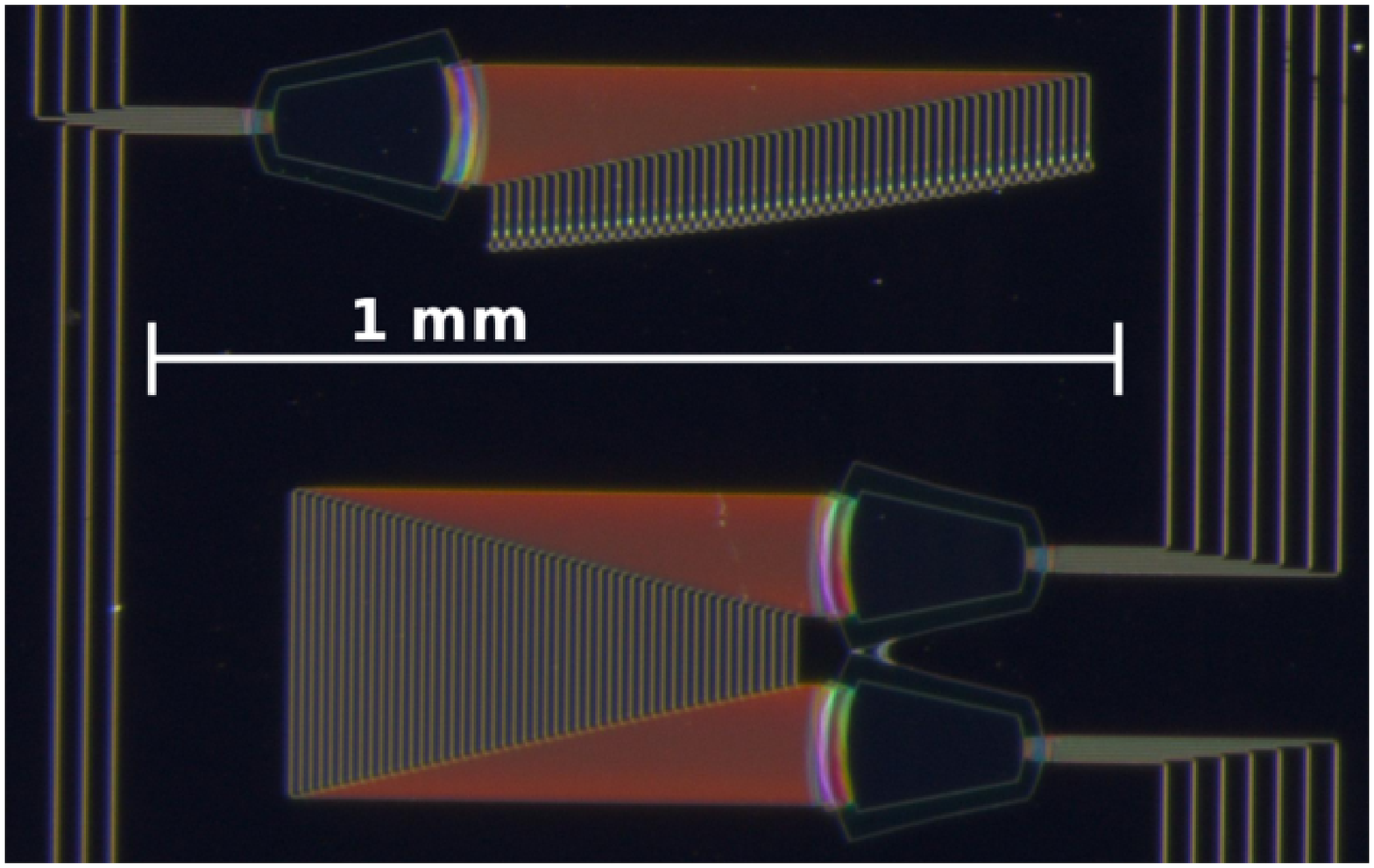}}}
  \subfigure[AWG spectra]{\resizebox*{0.43\textwidth}{!}{\includegraphics*{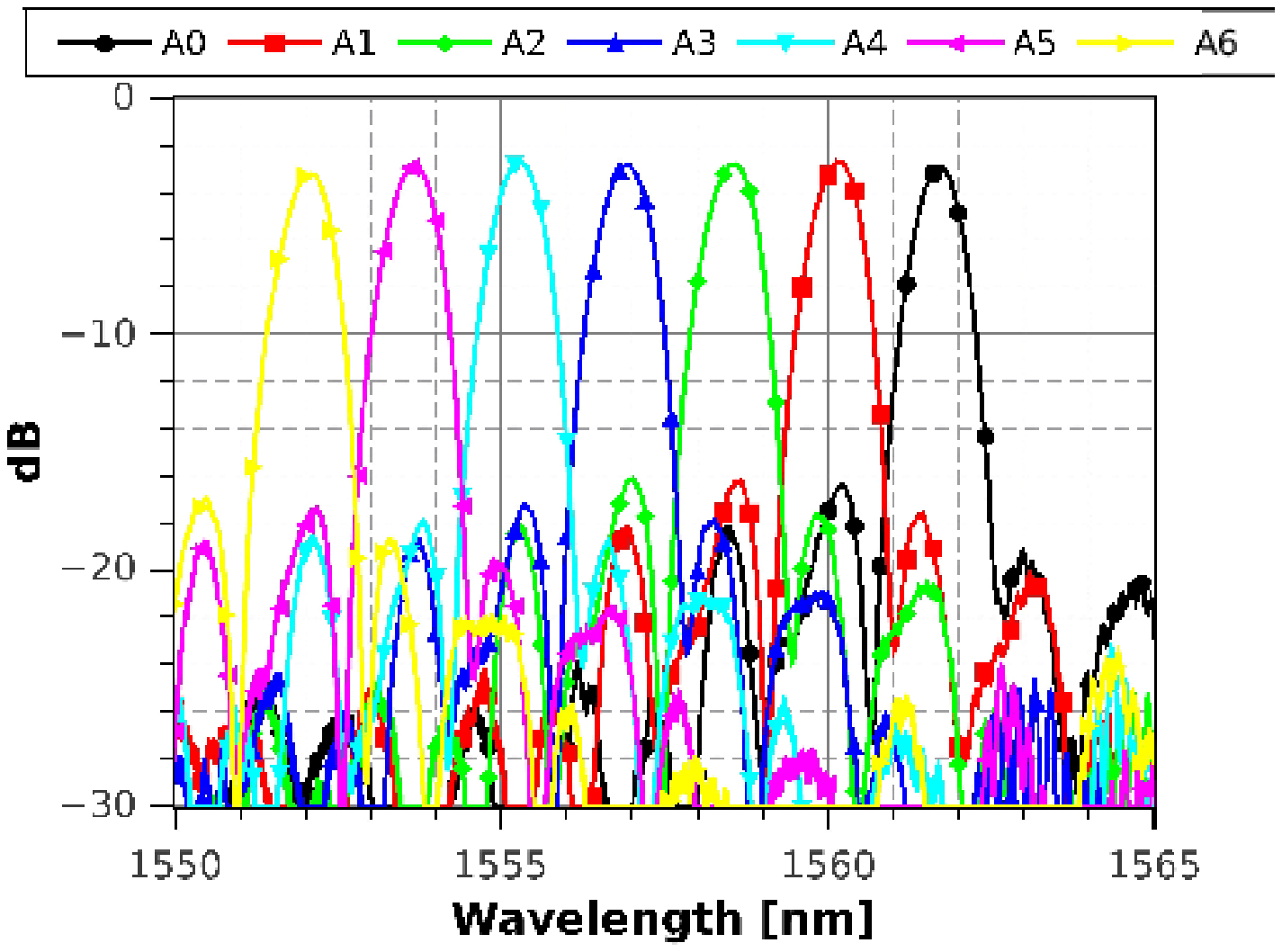}}}
  \subfigure[R-AWG spectra]{\resizebox*{0.43\textwidth}{!}{\includegraphics*{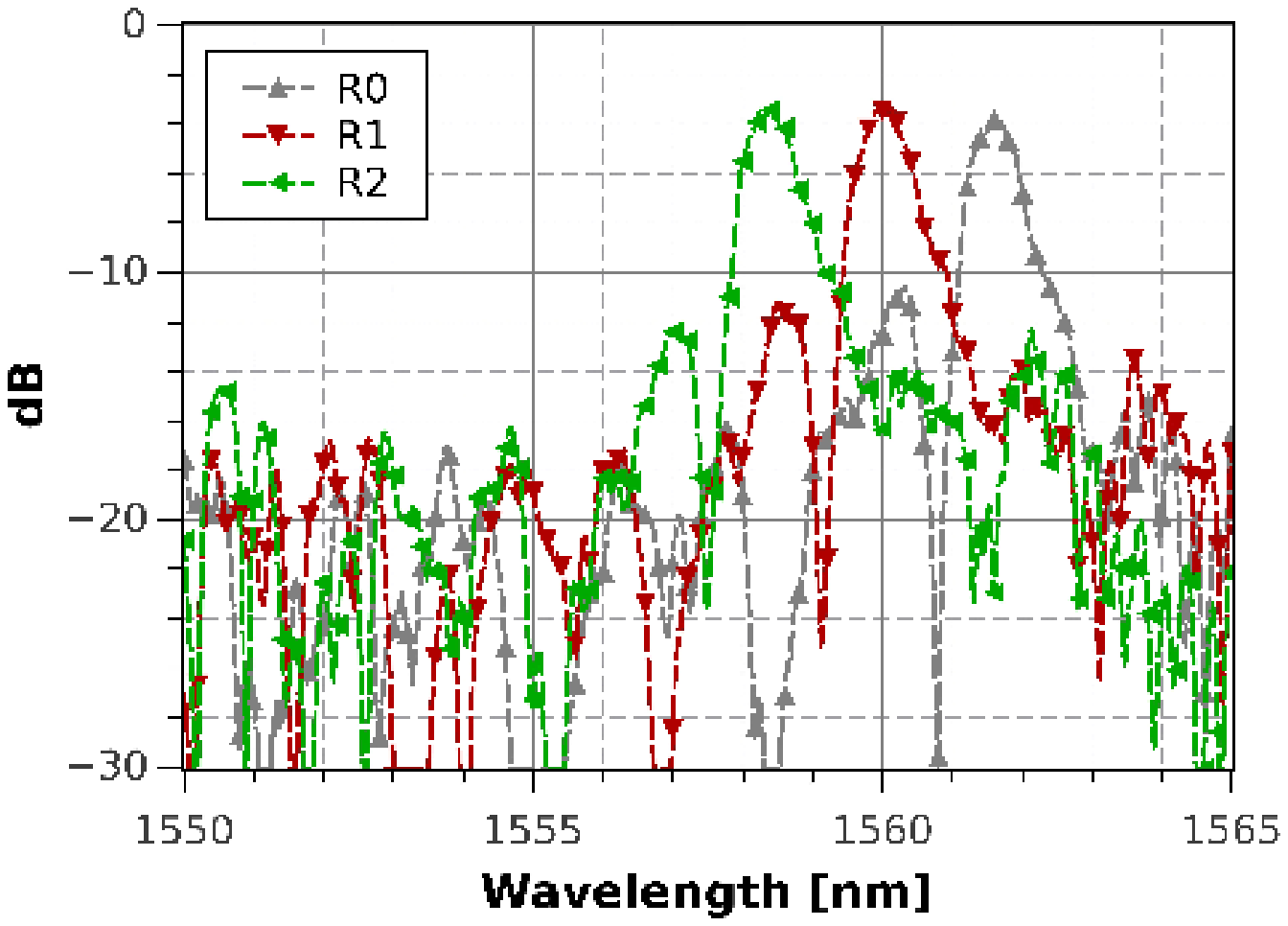}}}
  \subfigure[Comparison R-AWG vs AWG]{\resizebox*{0.43\textwidth}{!}{\includegraphics*{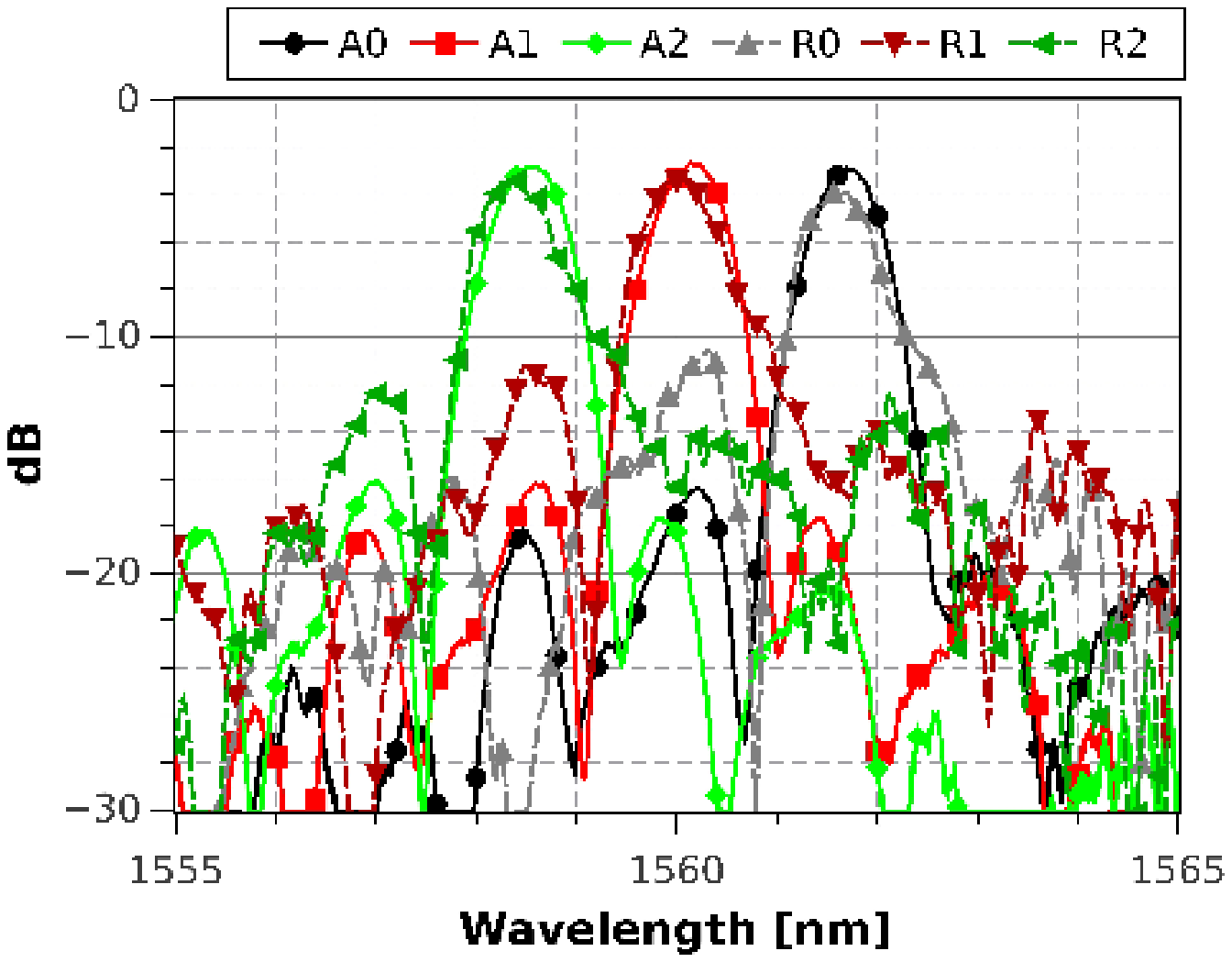}}}
  \vspace{-3mm}
  \caption{\label{fig:awg} Optical microscope image of the fabricated AWGs (a) and spectral traces, (b) regular AWG, (c) R-AWG and (d) comparison.}
}
\end{figure}

Following the model and methodology from \cite{gargallo_arxiv_rawg_model}\cite{pascual}, regular and reflective AWGs were designed having as target polarization TE, on a Silicon-on-Insulator (SOI) substrates consisting of a 3~$\mu$m thick buried oxide layer and a 220~nm thick Si layer, with no cladding. The effective indexes, calculated using a commercial software, are 2.67 in the arrayed waveguides ($n_{c}$) -waveguide with 0.8 $\mu$m to minimize phase errors, see \cite{bogaerts}- and 2.83 in the slab coupler ($n_{s}$). The R-AWG parameters are the following: the center wavelength is 1550 nm, using 7 channels with a spacing of 1.6~nm and a free spectral range (FSR) of 22.4~nm. The calculated focal length is 189.32~$\mu$m, the incremental length between AWs is 36.03~$\mu$m and the number of AWs is 49. The bend radius was set to 5 $\mu$m. The R-AWGs has a footprint of 350x950 $\mu$m$^2$ (width x height) following an orthogonal layout. The fabricated devices are shown in Fig.~\ref{fig:awg}-(a). Each waveguide in the R-AWG array is terminated by a SLR built with a 1x2 Multimode Interference coupler, with 50:50 splitting ratio for ideally full reflectivity. The input/output waveguides are equipped with focusing grating couplers (FGCs).

The waveguides are fabricated on the SOI substrates by Electron Beam Lithography (EBL) and dry etching in a two-step process. First, using hydrogen silsesquioxane negative tone resist in combination with a high contrast development process \cite{x} all device features are defined and fully etched to the buried oxide using an HBr-based ICP-RIE process \cite{y}. In a second step, a positive tone ZEP resist mask is carefully aligned to those features, exposed and used to define the shallow etched parts of the devices using a C4F8/SF6-based dry etching process. For both process steps a multi pass exposure approach is used to further reduce the sidewall roughness of the photonic device, hence minimizing scattering losses in those devices. Furthermore, special care is taken to guarantee accurate CD of all parts of the device by applying a very accurate proximity effect correction in combination with a well-balanced exposure dose \cite{z}.

For spectral characterization, a broadband source was employed in the range of 1525-1575~nm, and traces were recorded using a Optical Spectrum Analyzer with 10~pm resolution. All the traces were normalized with respect to a straight waveguide. The results are shown in Fig.~\ref{fig:awg}. Panel (b) shows the spectra for the seven channels of the AWG, from the central input. Peak insertion loss is approximately 3~dB. Note this value is subject to small variations in the performance of the FGCs (expected $\pm$0.4 dB). The highest side lobe level is 12~dB below the pass band maximum. Panel (c) shows the the spectra for the three inner channels from the central input, for the R-AWG. The other three channels were not designed to be measured, as they end in the same side of the chip. Finally, panel (d) shows the comparison of both AWG and R-AWG. Two main differences are clearly visible in the figure, between the AWG and the R-AWG. These can be appreciated in panel (d) comparing for instance traces A0 and R0. First, the shape of the pass band is slightly degraded towards longer wavelength, for the R-AWG, where broadening happens at 6~dB below maximum. Second, the side lobe level is increased by 4~dB in the R-AWG as compared to the AWG. Being the only difference between both devices the presence of SLRs, these degradations are likely to be due to phase/amplitude imperfections in the reflectors. 

\section{Conclusion and outlook}
In conclusion, we have reported the experimental demonstration of a SOI reflective Arrayed Waveguide Grating, with Sagnac mirrors and Gaussian spectral response. The performance of this first prototype is comparable to that of a regular twin AWG in the same die. Differences, likely due to dissimilarities between reflectors, are subject of current on-going research. The demonstration is the first step towards arbitrary pass band response R-AWGs employing SLRs with different reflection coefficients in each arm.

\section*{Acknowledgement}
The authors acknowledge the Spanish MICINN TEC2010-21337, MINECO TEC2013-42332-P, FEDER UPVOV 10-3E-492, FEDER UPVOV 08-3E-008 and FPI BES-2011-046100.


\end{document}